\documentclass[preprint,12pt]{elsarticle}

\usepackage{amsmath,amssymb}
\usepackage{graphicx}
\usepackage{hyperref}
\hypersetup{
    colorlinks=true,
    linkcolor=blue,
    citecolor=blue,
    urlcolor=blue
}
\usepackage[
    top=2.5cm,
    bottom=2.5cm,
    left=2cm,
    right=2cm
]{geometry}
\usepackage{float}

\usepackage[utf8]{inputenc}
\usepackage[T1]{fontenc}
\usepackage{ulem}
\usepackage{color}

\newcommand{\bs}{\boldsymbol}

\newcommand{\beq}{\begin{equation}}
\newcommand{\eeq}{\end{equation}}
\newcommand{\barr}{\begin{eqnarray}}
\newcommand{\earr}{\end{eqnarray}}
\newcommand{\mbf}{\mathbf}

\journal{Physics Letters A}

\begin{document}
	
	\begin{frontmatter}
		
		\title{Cherenkov radiation in isotropic chiral matter: the space-frequency domain}
		
		\author[ICN]{R. Mart\'inez von Dossow\corref{cor1}}
		\ead{ricardo.martinez@correo.nucleares.unam.mx}
		\cortext[cor1]{Corresponding author}
		
		\author[ITMO]{Eduardo Barredo-Alamilla}
		\ead{eduardo.barredo@metalab.ifmo.ru}
		
		\author[ITMO]{Maxim A. Gorlach}
		\ead{m.gorlach@metalab.ifmo.ru}
		
		\author[ICN]{Luis F. Urrutia}
		\ead{urrutia@nucleares.unam.mx}
		
		\address[ICN]{Instituto de Ciencias Nucleares, Universidad Nacional Aut\'onoma de M\'exico, 04510 Ciudad de M\'exico, M\'exico}
		\address[ITMO]{School of Physics and Engineering, ITMO University, 197101 Saint Petersburg, Russia}
		
		\begin{abstract}
The electromagnetic response of isotropic chiral matter, as described by Carroll–Field–Jackiw electrodynamics, arises in distinct physical contexts ranging from condensed matter systems to Lorentz-violating extensions of high-energy physics. Here, we  derive exact expressions for the circularly polarized electromagnetic fields that contribute independently to Cherenkov radiation in isotropic chiral matter. Each spectral energy distribution is gauge-invariant and positive, yielding radiation that emerges at a characteristic angle, akin to the standard case.  Furthermore, we identify specific frequency ranges that permit zero, one, or two Cherenkov cones for a given setup.  Remarkably, one sector of the model allows for the existence of threshold-free Cherenkov radiation arising from slowly-moving charges.
\end{abstract}
		
		
	\end{frontmatter}

\section{Introduction}

Since its discovery \cite{Cherenkov:1934ilx, 
 vavilov1934cr} and subsequent understanding in terms of Maxwell's equations \cite{Frank:1937fk}, Cherenkov radiation (CHR) has played a fundamental role in physics  ranging from Cherenkov
 detectors \cite{Ypsilantis:1993cp, E598:1974sol, TibetASgamma:2021tpz}, light sources \cite{adamo2009light, liu2012surface,liu2017integrated}, and more recently medical imaging \cite{hachadorian2020imaging, alexander2021color, shaffer2017utilizing},
and  photodynamic therapy \cite{wang2022cherenkov, kotagiri2015breaking, kamkaew2016cerenkov}. Current research in this area is very active and can be included in the more general topic of radiation engineered  via structured environments,  among which  we identify the following settings: two-dimensional materials, metamaterials, photonic crystals and external fields. For a detailed review see for example \cite{zrelov1970cherenkov, hu2021free}.

The fundamental restriction arising from the Cherenkov threshold, i. e. that the velocity of the charge must be larger than the velocity of light in the medium, requires high energy particles usually available only in high-energy accelerators. In particular, the energy range in  medical applications is of the order of MeV's
in comparison with the Gev-range required to identify   charged particles in accelerators. This has generated a significant line of research related to the production of Cherenkov radiation using charges with very low speeds compared to those required by the original threshold, i.e. threshold-free Cherenkov radiation \cite{liu2017integrated, zhang2022tunable, hu2020nonlocality,  gong2023interfacial}. 
 Another aspect that hinders the use of Cherenkov radiation in applications is the fact that both the radiation and the particles that produce it move forward, which makes it difficult to separate the Cherenkov cone. This has motivated research into reversed Cherenkov radiation  (RCHR) \cite{skryabin2017backward, genevet2015controlled, galyamin2009reversed},  which was theoretically proposed in \cite{ veselago1967electrodynamics} invoking materials with negative index of refraction, dubbed left-handed materials.  The construction of such metamaterials \cite{shelby2001experimental} has made possible  a number of observations  of this reversed Cherenkov radiation \cite{xi2009experimental,zhang2009flipping,lu2019generation,duan2017observation}. 
 Contrary to earlier assumptions that negative refractive index materials were necessary, recent research has demonstrated that RCHR can occur in natural materials with positive refractive indices. 
A first step was taken in Ref. \cite{Franca:2019twk} where the existence of RCHR was demonstrated when a particle is incident perpendicularly to the interface between vacuum and a topological insulator (generally a magnetoelectric medium) whose electromagnetic response is described by axion electrodynamics.

  This mechanism  can be characterized as an interfacial process in the  jargon suggested in Ref.  \cite{gong2023interfacial}.  On the other hand, in Ref. \cite{chen2025gain} the emergence of RCHR in a positive-index isotropic slab exploiting optical gain  is presented.
  
Carroll-Field-Jackiw (CFJ) electrodynamics \cite{Carroll:1989vb},
a specific case of    axion electrodynamics \cite{Sikivie:1983ip,Wilczek:1987mv}, has been extensively explored at least  in  two main contexts: high-energy physics models that break Lorentz symmetry \cite{Adam:2001ma,Kostelecky:2002ue,Lehnert:2004hq, Lehnert:2004be,Kaufhold:2005vj,Colladay:2016rmy, Schreck:2017isa, Lisboa-Santos:2023pwc, OConnor:2023izw},  and condensed matter physics models dealing with the electromagnetic response of magnetoelectric media, \cite{Franca:2019twk, Silva:2020dli, Silva:2021fzh, Franca:2021svc,Franca:2021irg, Barredo-Alamilla:2023xdt,Silva:2023ffk,Franca:2024fav}. Here we focus on the  electrodynamics of isotropic chiral matter, a specific instance of CFJ electrodynamics, which extends standard electrodynamics by incorporating an additional magnetic current proportional to $\sigma \mbf{B}$ in Ampere's law.

We stress that the term ``chiral'' here does not refer to optical activity in reciprocal bi-isotropic (Pasteur) media described by the constitutive relations $\mathbf{D} = \varepsilon \mathbf{E} + i \xi_\text{c} \mathbf{B}$ and $\mathbf{H} = \mu^{-1} \mathbf{B} + i \xi_\text{c} \mathbf{E}$~\cite{lindell1994electromagnetic,serdyukov2001electromagnetics}, which has been extensively studied both historically and in modern contexts~\cite{Fernandez-Corbaton2016,voronin2024chiral,Totful2024Nov}. Instead, we focus on media exhibiting chiral anomaly, where the macroscopic manifestation appears directly in Maxwell’s equations as a Lorentz-violating term that produces an effective magnetic current $\sigma \mathbf{B}$, without requiring explicit frequency dispersion or spatial nonlocality, though such extensions are possible. For monochromatic plane waves, isotropic CFJ electrodynamics admits an equivalent description in terms of effective chiral response with the identification $\sigma = 2 \xi_\text{c}\omega$. Standard chiral media require $\xi_\text{c} \to 0$ as $\omega \to 0$, whereas the magnetic conductivity $\sigma$ can persist in the static limit. Important differences include: (i) bi-isotropic chiral media are stable, while isotropic chiral matter may admit unstable modes~\cite{qiu2017electrodynamics}; (ii) while reciprocal bi-isotropic chiral media produce polarization rotation upon transmission but not upon reflection~\cite{jaggard1990recent}, isotropic CFJ-type response can induce nonreciprocal effects such as cross-polarized transmission and reflection at time-interfaces~\cite{2rrr-glyn}. The CFJ-electrodynamics framework effectively describes diverse systems such as Weyl semimetals~\cite{Zyuzin2012Apr,Zyuzin2012Sep}, quark–gluon plasmas~\cite{Kharzeev2008Oct}, and dark matter~\cite{Donghan2024Sep}. Recently, it has been shown that such kind of response can be realized in realistic  photonic systems whose magnetization rapidly varies in time 
\cite{2rrr-glyn}.

 We consider  the radiation produced by a charge moving at constant velocity across  isotropic chiral matter, a topic thoroughly examined in previous studies  \cite{Altschul:2014bba,Schober:2015rya, Altschul:2017xzx,DeCosta:2018nyf, Tuchin:2018sqe,Tuchin:2018mte,Tuchin:2020gtz, Hansen:2020irw,Hansen:2023wzp}. This problem is particularly challenging due to the lack of gauge invariance in both local energy density and local Poynting vector, unlike standard electrodynamics. Furthermore, the local energy density is not positive definite, and the existence of runaway modes in plane wave solutions raises concerns due to imaginary frequencies in the dispersion relations.
To address these issues, we adopt a more unbiased perspective, recognizing that plane waves are not the propagating normal modes in this electrodynamics. Instead, we emphasize the need to determine the actual normal modes to properly assess the problem. Given the uncommon features of isotropic chiral electrodynamics, doubts arise regarding its observable consequences, particularly those related to radiation. Some studies propose that instabilities might forbid "vacuum", i.e. $\epsilon=1, \sigma\neq 0$, Cherenkov radiation due to unexpected cancellations among positive and negative energy  modes \cite{Altschul:2014bba,Schober:2015rya}, while others report the existence of high-energy radiation in "vacuum" \cite{Tuchin:2018sqe,Hansen:2020irw}.

Motivated by this controversy,  but going one step further,  we present an alternative derivation for the general case $\epsilon \geq 1$, with exact results in the parameters $\sigma$ and $v$ yielding a description of CHR in a real chiral  material.  We circumvent the issue of imaginary frequencies by employing a standard method in electrodynamics: solving Maxwell's equations in the space-frequency domain \cite{Jackson:1998nia, schwinger2019classical}. This approach involves performing a time Fourier transform on all fields and operators, requiring only real measurable frequencies.
We closely follow the original method proposed in Ref. \cite{Frank:1937fk} to describe standard CHR.
The work is organized as follows: we solve the modified Maxwell's equations in the space-frequency domain, obtaining the system's normal modes that satisfy a distinct dispersion relation characterized by two polarization indices. The full solution is a superposition of these modes, with coefficients determined by boundary conditions at the sources and infinity. We derive the electromagnetic potentials necessary for calculating the Poynting vector and demonstrate that the potentially problematic contribution to the spectral energy distribution is gauge invariant. Causality is ensured by imposing outgoing propagating waves at infinity, determining the final dispersion relations and radiation conditions.
Our calculation reveals that the radiation is an independent superposition of two polarization modes, each with  separate  physical meaning and experimentally distinguishable. Notably, each spectral energy distribution is positive definite and characterized by a well-defined emission angle, resulting in a characteristic Cherenkov cone. We validate our results by comparing them with previous calculations, finding complete agreement. To this end, an expansion of the electromagnetic fields to second order in the parameter $\sigma$ is presented in the Appendix. The paper concludes with a summary and conclusions.

 \section{Maxwell's equations}
We consider the case of isotropic chiral matter
with permittivity $\epsilon$, permeability $\mu=1$ and magnetoelectric susceptibility $\sigma$ governed by 
  a restricted CFJ electrodynamics described by the following  Maxwell's equations in Gaussian units 
  \begin{eqnarray}
&&\boldsymbol{\nabla}\cdot\epsilon\mbf{E}=4\pi\bar{\rho}, \qquad \qquad  \boldsymbol{\nabla}\cdot \mbf{B}=0,  \label{MAXW1}
		\\
&&\boldsymbol{\nabla}\times\mbf{E}=-\frac{1}{c} \frac{\partial \mbf{B} }{\partial t}, 	\qquad \qquad 	\boldsymbol{\nabla}\times \mbf{B}=\frac{4\pi}{c}\mbf{J} + \frac{\epsilon}{c}\frac{\partial \mbf{E} }{\partial t} + \frac{\sigma}{c} \mbf{B},
\label{MAXW2}
\end{eqnarray}
where $c$ is the speed of light in vacuum. 
The speed of light in the medium is $c/n$ with $n=\sqrt{\epsilon}$ being  the index of refraction . In real materials permittivity and permeability exhibit frequency dependence. Here, we assume $\epsilon$ to be constant within a finite frequency range where this approximation holds, in order to isolate the effects of the magnetoelectric parameter $\sigma$.
The source is a charge $q$ moving along the $z$-axis with constant velocity $v$, with charge density $\bar{\rho}= q \, \delta(x)\delta(y)\delta(z-vt)$ and current $\mbf{J}=\bar{\rho} \, v \, \mbf{\hat{k}}$. As usual it is convenient to introduce 
the standard electromagnetic potentials $\mbf{A}$ and $\Phi$. 
Since we  aim at discussing radiation in this media, we write the corresponding local energy density ${U}$ and local energy flux ${\mbf S}$ 
\beq
    {U}= \frac{1}{8\pi}(\epsilon \mathbf{E}^2+\mathbf{B}^2) -\frac{1}{8\pi} \frac{\sigma}{c}\mathbf{A} \cdot \mathbf{B}, \qquad  \quad
    {\mathbf{S}} = \frac{c}{4\pi} \mathbf{E} \times \mathbf{B} + \frac{\sigma}{8\pi}\left(\mathbf{A} \times \mathbf{E} - \phi \mathbf{B}\right)\label{CONS}, 
    \eeq
satisfying the conservation equation $\partial_t {U}+\boldsymbol{\nabla}\cdot \mbf{ S}=0$ outside the sources. 
The terminology for these quantities, which can be  derived from the system's locally conserved energy-momentum tensor $T^{\mu\nu}$ as $U=T^{00}$ and $S^i= T^{i0}$,  highlights that the key physical global quantities in our case are the total energy $\int d^3 x \, T^{00}$ and the total radiated energy crossing a closed surface at infinity, both of  which must be gauge invariant. In fact, the local  energy-momentum tensor can be redefined as ${\tilde T}^{\mu\nu}=T^{\mu\nu}+ \partial_\rho \, S^{\mu\rho\sigma}$ with $S^{\mu\rho\sigma}=- S^{\rho \mu\sigma}$. This property guarantees $\partial_\mu {\tilde T}^{\mu\nu}=0$ as well as the same integrated quantities obtained from $T^{\mu\nu}$. 
Similar energy conservation laws can be constructed for bi-isotropic chiral media, although they hold rigorously only for time-harmonic excitations.
The expressions (\ref{CONS}) present two drawbacks: (i) the local  energy density  is not positive definite  and (ii) the above quantities are not gauge invariant. 
However,  we deal with these issues after obtaining the explicit solutions for the electromagnetic fields and potentials.
To this end  we find it convenient to work in cylindrical coordinates $\rho, \, \phi, \, z$ and also in the space-frequency domain  where the sources are 
$
{\bar \rho}(\rho, z, \omega) = \frac{q}{2\pi v \rho} \delta(\rho) e^{i\frac{ \omega}{v}z}, J_x =J_y=0,   
J_z= q v \bar{\rho}
$, 
as appropriate to the axial symmetry of the system.
 The form
of the sources in the space-time domain dictates that the general structure of the fields, illustrated here in terms of the electric field, which is 
$
\mbf{E}(\mbf{x},t)= \mbf{E}(x,y,z-vt).
$
This yields de Fourier transform in time
$
\mbf{E}(\mbf{x},\omega)=\int_{-\infty}^{+\infty} dt \,e^{i\omega t} \, 
\mbf{E}(\mbf{x},t)=e^{i kz} \,\mbf{E}(\rho,k), $ with  $k=\frac{\omega}{v}$.
We deal with Maxwell's equations (\ref{MAXW1}) and (\ref{MAXW2}) in this Fourier space,  where we  have $
\partial_t= -i\omega$ and  $\partial_z=+ik$.
In the following we take advantage of the factorizations of the fields in the form 
$\mbf{V}(\mbf{x}, \omega)=e^{i k z}\,\mbf{V}(\rho, \omega)$ by displaying  only  the contribution $\mbf{V}(\rho, \omega)$, unless confusion arises. 

In  obvious notation for the components of the fields in cylindrical coordinates we find the following results. The magnetic field components are given in terms of the electric field via the Faraday's law
\begin{equation}B_\rho= -\frac{c}{v} E_\phi, \qquad  B_z=-\frac{ic}{k \rho v}\partial_\rho(\rho E_\phi), \qquad  B_\phi=\frac{c}{v}E_\rho +\frac{ic}{k v}\partial_\rho E_z. 
\label{B}
\end{equation}
The components $E_\rho$ and $E_\phi$ get mixed according to 
\barr
	&&\Big(\rho^2 \partial^2_\rho  + \rho \partial_\rho -\left(1+\rho^2\alpha^2\right)\Big)E_\phi = -\dfrac{in^2 \omega^2 \sigma}{kc^3}\rho^2 E_\rho, \label{EPHI1} \\
	&& \Big(\rho^2 \partial^2_\rho + \rho \partial_\rho  -(1+\rho^2\gamma^2) \Big)E_\rho = \dfrac{ik\sigma}{c}\rho^2 E_\phi,
	\label{ERRO1}
\earr
where we introduce the notation 
\begin{equation}
	\alpha^2=\gamma^2-\frac{\sigma^2}{c^2}, \qquad 
	\gamma^2=k^2 \Big(1-\frac{n^2 v^2}{c^2}\Big).
	\label{alpha2gamma2}
\end{equation}
From the Gauss law, the remaining component of the electric field is
\beq
E_z= \frac{i}{k}\frac{1}{\rho}\partial_\rho(\rho E_\rho).
\label{EZ1}
\eeq
The operators in the left hand side of Eqs. 
(\ref{EPHI1}) and (\ref{ERRO1}) are identified as those corresponding to a modified Bessel function of order one, which motivates us to take $E_\phi$ and $E_\rho$ as linear combinations of $K_1$ and $I_1$ \cite{abramowitz1965handbook}. The boundary conditions
at $\rho \to 0$ as well as $\rho \to \infty$
forces us to include $K_1$ as part or the solution and to reject the contribution of $I_1$.Then we take
\beq
E_\phi(\rho)= X K_1(Q\rho), \qquad E_\rho(\rho)= Y K_1(Q\rho).
\eeq
Substituting in Eqs. (\ref{EPHI1}) and (\ref{ERRO1}) and using the relations among the derivatives of the modified Bessel functions  we obtain a matrix equation 
which zero-determinat condition provides the dispersion relation 
\begin{eqnarray}
Q^4-(\alpha^2+\gamma^2)Q^2+\alpha^2\gamma^2-\frac{n^2 \omega^2 \sigma^2}{c^4}&=&0,
\end{eqnarray}
yielding  the allowed momenta $Q$ of the solution. This is a quartic equation with solutions
 \begin{equation}
	Q^2_\nu(\omega)=\frac{\omega^2}{v^2}-\frac{n^2 \omega^2}{c^2}-\frac{\sigma^2}{2c^2}+\nu \frac{\sigma}{2c^2} \sqrt{\sigma^2+4n^2 \omega^2}, \quad \nu= \pm,
	\label{DISPREL}
\end{equation}
where we used $k=\omega/v$. In our conventions, the modes $\nu= + \, (\nu=-) $ describe left (right) circular polarization in the plane perpendicular to the wave vector. 
The zero-determinat condition yields also  the relation 
$
Y_\nu=i \Omega_\nu \, X_\nu$ with $ \Omega_\nu=
{k \sigma}/{(Q^2_\nu-\gamma^2)c}.
$
In this way the electric field is presented as
\begin{equation}
 E_\phi=\sum_{\nu=\pm } X_\nu \, K_1(Q_\nu \rho), \qquad E_\rho= i\sum_{\nu=\pm }\Omega_\nu\, X_\nu \, K_1(Q_\nu \rho), \qquad E_z=\frac{1}{k}\sum_{\nu=\pm }\Omega_\nu\, Q_\nu\, X_\nu \, K_0(Q_\nu \rho),
\label{TOTALE}
\end{equation}
where the expression for $E_z$ is obtained from Eq. (\ref{EZ1}).  Equations  (\ref{B}) for  the magnetic field give
\begin{eqnarray}
	&& \hspace{1.3cm} B_\phi=i\frac{c}{v}\sum_{\nu=\pm } \left( 1 -  \frac{Q^2_\nu}{k^2} \right) \,X_\nu \Omega_\nu K_1(Q_\nu \rho), \nonumber \\
	&&  B_\rho=-\frac{c}{v}\sum_{\nu= \pm }X_\nu K_1(Q_\nu \rho), \quad B_z = i\frac{ c}{\omega}\sum_{\nu=\pm } Q_\nu X_\nu  K_0(Q_\nu\rho). 
	\label{TOTALB}
\end{eqnarray}

\begin{table*}[!ht]
	\scalebox{0.87}{
	\renewcommand{\arraystretch}{2}
	\setlength{\tabcolsep}{10pt}
	\begin{tabular}{c c c c}
		\hline \hline
		\textbf{Field} & \textbf{Function of } $(\mathbf{x},t)$ & \textbf{Function of } $(\mathbf{x},\omega)$ & \textbf{Ref.} \\
		\hline
		$\mathbf{B}^{(0,1)}$ 
		& $ \displaystyle qv \frac{\sin\theta}{R^{2}}\, \mathbf{\hat{\bs{\phi}}}$ 
		& $ \displaystyle 2qk\, e^{ikz} K_{1}(k\rho)\, \mathbf{\hat{\bs{\phi}}}$ 
		& \cite{Altschul:2014bba} \\
		\hline
		$\mathbf{B}^{(1,1)}$ 
		& $\displaystyle \sigma \frac{qv}{2R} \left(2\cos \theta\, \mathbf{\hat{\bs{r}}} - \sin \theta\, \mathbf{\hat{\bs{\theta}}} \right)$ 
		& $\displaystyle
		\left.
		\begin{aligned}
		\sigma q\, e^{ikz} \Big( & \left[2K_{0}(k\rho) - (k\rho) K_{1}(k\rho) \right]\, \mathbf{\hat{k}} \\
		& -i (k\rho) K_0(k\rho)\hat{ \bs{\rho}}\Big)
		\end{aligned}
		\right.
		$
		& \cite{Altschul:2014bba} \\
		\hline
		$\mathbf{B}^{(2,1)}$ 
		& $\displaystyle \sigma^{2} \frac{qv}{2} \sin \theta\, \mathbf{\hat{\bs{\phi}}}$ 
		& $\displaystyle \sigma^{2} q\rho\, e^{ikz} K_{0}(k\rho)\, \mathbf{\hat{\bs{\phi}}}$ 
		& \cite{Schober:2015rya} \\
		\hline
		$\mathbf{E}^{(2,2)}$ 
		& $\displaystyle -\sigma^{2} \frac{qv}{4} \left[ \left(\frac{3}{2} \cos^{2} \theta - \frac{1}{2} \right) \mathbf{\hat{r}} - \sin \theta \cos \theta\, \mathbf{\hat{\bs{\theta}}} \right]$ 
		& $\displaystyle
		\left.
		\begin{aligned}
		\sigma^2 q\, e^{ikz}\,\frac{\rho}{4}\Big( & (\omega \rho) K_1(k\rho)\, \hat{ \bs{\rho}} \\
		& +    i v [2K_1(k\rho)-(k \rho)K_0(k \rho)]\, \hat{\mathbf{k}} \Big)
		\end{aligned}
		\right.
		$
		& \cite{Schober:2015rya} \\
\hline		\hline
	\end{tabular}
	}
	\caption{The fields in the space-time domain  are written in spherical coordinates with $\mbf{R}=\mbf{x}- \mbf{v} t, \, R=|\mbf{R}|, \, \cos\theta=(z-vt)/R, \, \sin\theta=\sqrt{x^2+ y^2}/R$. The fields in the space-frequency domain are in cylindrical coordinates following  the conventions in the manuscript. The parameter ``k'' in the cited references corresponds to $\sigma/2$  in our conventions. Here we set $c=1$.  }
	\label{tab:campos}
\end{table*}
 Finally, imposing  the boundary conditions at $\rho \rightarrow 0$  we obtain  the  
solutions
	$X_+ = -\frac{i\omega}{c}\, \Gamma \,Q_+, \quad  X_- = \frac{i\omega}{c}\, \Gamma\, Q_-,  
$
with $\Gamma={2q}/{\sqrt{\sigma^2+4n^2 \omega^2}}$.
The final expressions for the electric and magnetic fields are obtained by substituting these coefficients in Eqs. (\ref{TOTALE})  and (\ref{TOTALB}).
Let us recall that the full coordinate dependence of the above fields is obtained after  multiplying by $e^{ikz}$.

{This approach extends the iterative method proposed in Refs. \cite{Altschul:2014bba,Schober:2015rya}, which we validate by successfully comparing explicit iterations from those references with a series expansion in $\sigma$ of our exact results. We expand our total electric and magnetic fields up to second order in $\sigma$, with the details shown in the Appendix. Next, we examine the fields listed in the first column of Table \ref{tab:campos}, recalling their expressions given in the second column. These fields were previously calculated  in the space-time domain  in the references cited in the last column. Next, we present our calculation of the time Fourier transforms of these fields in the third column  and successfully compare the results with those reported in the Appendix.

To conclude the calculation of the required electromagnetic  components we deal with the potentials entering the Poynting vector. This is a simple task in the space-frequency domain once we know the electromagnetic fields.  We work in  the Coulomb  gauge 
$
\boldsymbol{\nabla}\cdot \mbf{A}=0, \, \Phi=0
$
outside the sources, where ${\mbf A}=-ic{\mbf E}/\omega$.

\section{The radiation zone}
\label{RADAPP}

 The spectral energy distribution in radiation arises from the total energy $E$ flowing across a closed surface $\cal S$ at infinity given by $E= \int_{-\infty}^{+\infty} dt \oint_{\cal S} d{ \cal S} \,  \hat{\boldsymbol{n}}\cdot 
{\mbf{S}}(\mbf{x}, t)$. Let us focus on the conflicting term of the Poynting  vector  given in Eq. (\ref{CONS}),  which depends  on the potentials. Performing the  gauge transformation $\delta_G \mbf{A}=\bs{\nabla} \delta \Lambda, \, \delta_G \Phi=-\frac{1}{c}\partial_t \delta \Lambda $ on its contribution    $\tilde E$ to the total radiated energy yields
\beq
\delta_G \tilde{E}=\frac{\sigma}{8\pi} \int_{-\infty}^{+\infty} dt \oint_{\cal S} \Big((\boldsymbol{\nabla} \delta \Lambda) \times \mbf{E} + \frac{1}{c} (\partial_t \delta \Lambda)\, \mbf{B}\Big)\cdot \hat{\mbf{n}}\,  d {\cal{S}}
\label{GT1}
\eeq 
Assuming $(\delta \Lambda  \mbf{B})$ vanishes at 
$t=\pm \infty $, we integrate by parts in time the 
second term in the right hand side of  (\ref{GT1}) and use Faraday's law together with Gauss theorem, obtaining
\barr
\delta_G \tilde{E}&=&\frac{\sigma}{8\pi} \int_{-\infty}^{+\infty} dt \oint_{\cal S} \Big((\boldsymbol{\nabla} \delta \Lambda) \times \mbf{E} + \delta \Lambda \boldsymbol{\nabla}\times \mbf{E} \Big)\cdot \hat{\mbf{n}} d {\cal{S}}  
=\frac{\sigma}{8\pi} \int_{-\infty}^{+\infty} dt \int_V d^3 x \, \boldsymbol{\nabla} \cdot \mbf{V}, 
\label{GT22}
\earr 
where we denoted by $\mbf{V}$ the vector in round brackets in the second term of Eq. (\ref{GT22}). Using standard vector identities it is a direct matter to show that  $\boldsymbol{\nabla} \cdot \mbf{V}=0$,  which proves  the gauge invariance of $\tilde E$.

In our case we  consider the surface of an infinite cylinder with axis along the $z$-direction and radius $\rho \rightarrow \infty$. Starting from  standard electrodynamics, i. e $\sigma=0$, where we have $ E =  \rho \int_{0}^{+\infty} \frac{d\omega}{2\pi} \int_{-\infty}^{+\infty} dz\, \hat{\boldsymbol{\rho}}\cdot \, {\rm Re}( c \,  \mbf{E}^*\times \mbf{B})$, we need to include the additional contributions of the  Poynting vector in (\ref{CONS}). After doing this we read  the spectral distribution of the total radiated energy per unit lenght ${\cal E}= \frac{d^2 E}{d \omega dz}$
 \beq
{\cal E}=\lim_{\rho \to \infty} \frac{\rho}{2\pi}  \mathrm{Re} \Big[ c E^*_{ \phi} B_z -  c E^*_zB_\phi 
+ \frac{\sigma}{2}(A_\phi E^*_z-A_z E^*_\phi -\Phi\, B^*_\rho) \Big], 
\label{FINALSED}
\eeq
which is finally written in the space-frequency domain, as a function of the fields (\ref{TOTALE}) and  (\ref{TOTALB}).

To investigate the conditions for non-zero radiation, we study the fields in the asymptotic region $\rho \rightarrow \infty$. In particular, our aim is to determine the range of frequencies in which  radiation is allowed for a given set of parameters $\sigma, n , v$, and also which polarizations $\nu$ contribute.
All the involved fields are functions of the modified Bessel functions $K_{0,1}(Q\rho)$   having the asymptotic behavior
$
\lim_{\rho \to \infty}K_{0,1}(Q_\nu\rho)=
\sqrt{\frac{\pi}{2 Q_\nu \rho}}e^{-Q_\nu \rho},
$
with $Q_\nu^2$ given by Eq. (\ref{DISPREL}).
At this stage, we impose causality by requiring outgoing waves at the cylindrical surface as $\rho \to \infty$.
This demands  to choose 
$
Q_\nu=-i{\cal Q}_\nu,
$
 with ${\cal Q}_\nu $ real and positive. Going back to the dispersion relation (\ref{DISPREL}) we factor a minus sign and choose ${\cal Q}_\nu= \sqrt{q_\nu}$, with 
\begin{equation}
	q_\nu\equiv \Big(\frac{n^2 \omega^2}{c^2}+\frac{\sigma^2}{2c^2}-\nu \frac{\sigma}{2c^2} \sqrt{\sigma^2+4n^2 \omega^2}\Big) -\frac{\omega^2}{v^2} \geq 0.	
	\label{POSDEF1}
\end{equation}
The relation $q_\nu \geq 0$ determines the conditions under which a particular polarization mode $\nu$ contributes to the radiation.

Prior to simplifying the condition (\ref{POSDEF1}) under which radiation occurs, we first determine the contribution of the two polarization modes $\nu=\pm 1 $ to the spectral energy distribution. The main objective is to examine potential interference between these components when computing the necessary quadratic products in the fields of Eq. (\ref{FINALSED}). Calling ${\cal E}=\frac{d^2 E}{d\omega dz}$ we expect in general
$
{\cal E}={\cal E}_+ + {\cal E}_- + {\cal E}_{(+,-) } 
$,
where each ${\cal E}_{\pm}$ is given by the expression (\ref{FINALSED}), substituting  the fields and potentials with polarization $\pm$, respectively.                           The term ${\cal E}_{(+,-) }$ describes the interference between the polarization modes. These contributions are directly calculated using the asymptotic expressions for the fields. Surprisingly, the interference term yield exactly zero, showing that the spectral energy distribution is  simply presented as a sum  of the two polarization modes. Our proof of gauge invariance at the beginning of this section is separately valid for each polarization, since the above cancellation shows that effectively we can write   $\mbf{S}=\mbf{S}_+ + \mbf{S}_- $. This means that the radiation produced by  each polarization has  an independent  physical contribution to the spectral energy distribution, which prevent us to look at the physical spectrum only  as the sum of the two contributions. As we will see in the following, each polarization is characterized by a definite
angle of emission, so that they are experimentally distinguishable by sweeping the observation angle in such a way that one cone, two cones, or no cone can appear.  
The calculation of ${\cal E}_\nu$ is simple but tedious, yielding the result
\begin{equation}
	{\cal E}_\nu = \frac{ q^2 \omega}{2c^2}\left( 1-\frac{1}{n^2 \beta^2}-\nu\frac{\sigma}{\sqrt{\sigma^2+4n^2 \omega^2}}\left(1+\frac{1}{n^2 \beta^2}\right)\right) \equiv \frac{q^2 \, \omega}{c^2} \, \Omega,
	\label{ELAMBDA2}
\end{equation}
where $\beta=v/c$ is the ratio of the velocity of the moving  charge over the velocity of the light in vacuum. 
This can be presented in the alternative form 
\begin{equation}
{\cal E}_\nu=\frac{ q^2 \omega}{2 c^2} \frac{{\cal F}_\nu}{\sqrt{1+\frac{ \Sigma^2}{4}}}  
\left[1-\frac{1}{ n^2 \beta^2 {{\cal F}_\nu}^2}\right],
\label{ELAMBDA}
\end{equation} 
which will prove useful in investigating the positivity properties of ${\cal E}_\nu$.  We introduce  the notation  $\Sigma = \frac{\sigma}{n \omega} > 0$ together with
\beq
    {\cal F}_\nu= \sqrt{1+\frac{\Sigma^2}{4}}-\nu\frac{\Sigma}{2}= 
    \frac{1}{2 n \omega}\Big(\sqrt{\sigma^2+ 4 n^2 \omega^2} -\nu \sigma\Big), 
    \label{FNU}
\eeq
    where  ${\cal F}_\nu$,  being of the form $\sqrt{1+ z^2} \pm z$ with $z$ real, is always positive.

 Now we calculate  the Cherenkov angles at which radiation is emitted. In the radiation zone, the fields in each polarization are proportional to $e^{i {{\cal Q}_\nu \rho}+ ik z}$ so that the wave vector is 
$
 {\mbf K}_\nu= {\cal Q}_\nu \, \hat{\bs{\rho}} +k \hat{\bs{k}},
$
 with 
 \beq
 |{\mbf K}_\nu|=\sqrt{{\cal Q}^2_\nu+ k^2}
 = \frac{1}{2c}\sqrt{\Big(\sqrt{\sigma^2+4n^2 \omega^2}-\nu \sigma\Big)^2} =
 \frac{n \omega}{c} {\cal F}_\nu,
 \eeq
     where we use  ${\cal Q}^2_\nu=q_\nu$ from Eq. (\ref{POSDEF1}),  together with  ${\cal F}_\nu > 0$. We obtain  the angle of the emitted radiation with respect to $z$-axis, $\Theta_{\nu}$, as 
 \beq
 \cos \Theta_\nu=\frac{k}
 {|{\mbf K}_\nu|}= \frac{1}{n \beta {\cal F}_\nu},
 \label{THETAC}
 \eeq 
 requiring 
 $
 \beta  \geq {1}/ {n {\cal F}_\nu}, 
 $
 which is the condition for the charge to radiate in the polarization mode $\nu$. 

Next we  confront the second issue in isotropic chiral  electrodynamics: the question whether or not  the total radiated energy is positive definite. Even thought the general expression for the local energy density in (\ref{CONS}) does not guarantee positivity, it does not rule  it out either, so that it is necessary to verify what happens in each particular case.   
Equation (\ref{THETAC}) allows us to rewrite  (\ref{ELAMBDA}) as 
\begin{equation}
{\cal E}_\nu=\frac{ q^2 \omega}{2 c^2} \frac{{\cal F}_\nu}{\sqrt{1+\frac{ \Sigma^2}{4}} }
\sin^2 \Theta_\nu,
\label{ELAMBDAtheta}
\end{equation} 
clearly showing that  the total radiated energy for each polarization is positive, in spite of the uncertaintities stemming from the first  equation  in (\ref{CONS}). Notice that in the limit $\sigma \to 0$, both  angles coalesce into the standard Cherenkov $\Theta_{\rm s}$ angle satisfying $\cos \Theta_{\rm s}= 1/(\beta n)$. 
 
Given that each polarization has been shown to possess a physically meaningful and experimentally verifiable interpretation, we now explore in more detail the conditions for radiation to occur, given in Eq. (\ref{POSDEF1}).
A further simplification  arises since the term in round brackets is the perfect square $(\sqrt{\sigma^2 +4n^2 \omega^2}-\nu \sigma)^2/4c^2$, where  $(\sqrt{\sigma^2 +4n^2 \omega^2}-\nu \sigma)=2n \omega {\cal F}_\nu$ is positive definite. This allows to present Eq. (\ref{POSDEF1}) as 
\beq
\frac{1}{2c}\sqrt{\sigma^2 +4n^2 \omega^2}\geq \frac{\omega}{v} + \frac{\nu \sigma}{2c}.
\label{COND3}
\eeq 
Let us emphasize that this condition is the same  as that required  for the existence of the radiation angles $\Theta_\nu$ defined  in Eq. (\ref{THETAC}).  
A further  rewriting of Eq. (\ref{COND3}) leads to the final condition for radiation
\beq
\omega (n^2 \beta^2-1)\geq \nu \sigma \beta,
\label{CONDRADFIN}
\eeq
where  $\omega \geq 0$. 

The signs in the inequality (\ref{CONDRADFIN}) are properly taken into account by considering the following  cases,  which define the contribution of the polarization modes:

(i) When $n=1$ and   $\beta^2-1<0$  we have  the propagation of the polarization $\nu=-$ only, provided  $0 < \omega < \sigma \beta/(1-\beta^2)$. Observe that for $\sigma=0$ the allowed window for $\omega$ closes and  there is no radiation at all, as expected.  

 (ii) When $n >1 $ and  $(n^2 \beta^2-1) >0$ we have that $v>c/n$, i. e. the velocity of the charge is larger that the velocity of light in the medium. Then the inequality  (\ref{CONDRADFIN}) yields
 \beq
 \omega > \frac{\nu \sigma \beta}{(n^2 \beta^2-1)}\equiv \nu \omega_0.
 \label{CONDVGEQVL}
 \eeq
 Thus the channel  $\nu=+$ opens up when $\omega > \omega_0$ while  it is  always open  for $\nu=-$.  In other words, we always  have one cone, $\nu=-$, together with a second cone provided $\omega>\omega_0$.
The correlation between the radiation angles and different  particle velocities is shown in the left panel of Fig. \ref{FIG12} as a function  of the frequency, which allows us to identify the onset of the mode $\nu=+$. As the velocity increases, the angles of both modes converge to the standard Cherenkov angle, resulting in reduced angular sensitivity. However, this sensitivity can be adjusted by varying either $n$ or $\sigma$. Notably, increasing $n$ while keeping $\beta$ and $\sigma$ fixed decreases the sensitivity. In contrast, for fixed $\beta$ and $n$, increasing $\sigma$ enhances the sensitivity. The latter situation is shown in the right panel of Fig. \ref{FIG12}.  
 \begin{figure}[h!]
	\centering
	\includegraphics[scale=0.8]{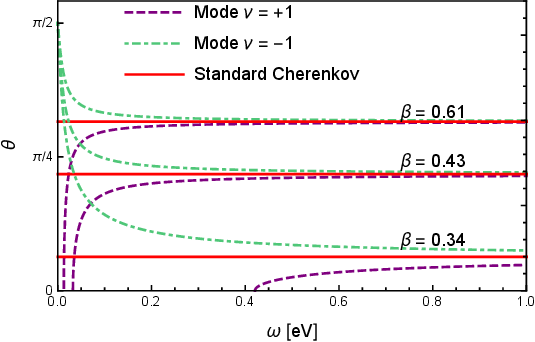}
	\hspace{0.2cm}
	\includegraphics[scale=0.8]{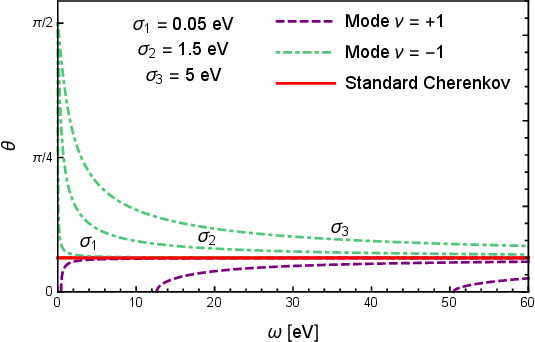}
	\caption{Left panel: Emission angles as a function of the frequency, for different particle velocities. The parameters are $\sigma=0.05 {\rm eV}$ and $n=3 \, \, (\beta >1/3$). Right panel: Emission angles as a function of the frequency, for different values of $\sigma$, with $\beta=0.34$ and $n=3$.}
	\label{FIG12}
\end{figure}
 
From Eq. (\ref{ELAMBDA2}), it is interesting to observe that when both polarization channels are open   the total emitted  energy  
\beq
{\cal E}={\cal E}_+ + {\cal E}_-= \frac{q^2 \omega}{c^2}\Big( 1-\frac{1}{n^2\beta^2} \Big)
\label{ELAMBDA4}
\eeq
turns out to be  independent of $\sigma$,
and coincides with that of the standard CHR in an isotropic media with the same refraction index. However, each separate mode  is radiating  at a given  angle with  an explicit dependence on this parameter.  For $\sigma=0.05$ eV, $n=3$  and $\beta= 0.43$,  ${\cal E}$ results of the order of 
    $10^{-32}\,$-$\,10^{-31}$ [J-s/m] in the frequency range $0 < \omega  < 0.1$ eV. 
 It is a consistency check to verify the limit
 as $\sigma=0$. This yields $\omega_0=0$ and $\Theta_+ = \Theta_-= \Theta_s$. Then, the total radiated energy flows through the standard Cherenkov angle $\Theta_s$ and the spectral distribution  is given by ${\cal E}$ in Eq. (\ref{ELAMBDA4}), which reproduces  the standard result for CHR when $n>1$. 
 
(iii) When $n>1$ and  $(1-n^2 \beta^2) >0$, we have $v<c/n $ and (\ref{CONDRADFIN})  demands
 \beq
 0< \omega< -\frac{\nu \sigma \beta}{(1-n^2 \beta^2)}=-\nu |\omega_0|.
 \eeq
This can  only be satisfied for $\nu=-$, and $\sigma\neq 0$. 
This sector  of the model allows for the study of the most sought-after threshold-free CHR  \cite{liu2017integrated, zhang2022tunable, hu2020nonlocality,  gong2023interfacial}. A key figure of merit here is the photon extraction efficiency per unit length for the channel $\nu=-1$, defined as ${\tilde \eta}_{-1}= d \eta_{-1}/dz={\cal E}_{-1}/(\hbar \omega \, E_{\rm ch})$  \cite{gong2023interfacial, Chen:2022qlr}.   This  can be interpreted as the total number of photons emitted per unit length and per unit frequency divided by the  kinetic energy of the charge $E_{\rm ch}= m_{\rm ch} c^2 (1-\gamma)$. Here $\gamma$ is the standard Lorentz  factor. The result is 
\beq
{\tilde \eta}_{-1}= f \frac{1}{(\gamma-1)} \Omega, \qquad  f=\frac{q^2}{c^4\, \hbar \, m_{\rm ch}},
\label{FEE1}
\eeq
where $\Omega$ is defined in Eq. (\ref{ELAMBDA2}). 

 For an  electron we  illustrate the main features of ${\tilde \eta}_{-1} $ by focusing in a Weyl semimetal as a representative of isotropic chiral matter \cite{borisenko2019time,lv2015observation,grassano2020influence,zu2021comprehensive}.  We choose the following order of magnitude values  $\sigma=0.05$ eV and  $n=3$, yielding $\beta_{\rm max}= 1/3$ as the maximum velocity  . The constant factor is  $f=296.8 \,  [{\rm Watt \, m}]^{-1}$. 
We find that  $\Omega$  is a slowly varying function  in the interval  $\omega < \omega_0$, which goes from one to zero in the allowed region,  as shown 
in Fig. \ref{FIG11}. Let us observe that  for small velocities, we have a smaller frequency range and the curve grows more rapidly to its maximum value of $1$.

\begin{figure}[h!]
	\centering
\includegraphics{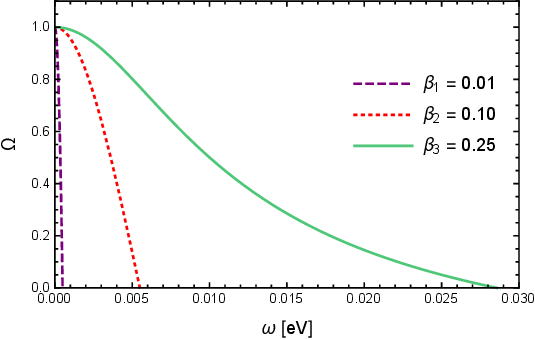}
	\caption{The function  $\Omega(\omega)$ for the charge velocities $\beta_1=0.01$, $\beta_2=0.10$ and $\beta_3=0.25$.}
	\label{FIG11}
\end{figure}

In other words,  the strongest variations of ${\tilde \eta}_{-1}$ arise from the amplifying  factor $1/(\gamma-1)$ as schematically shown in Table \ref{T1}.  However, let us notice  that as this factor grows, the allowed frequency range diminishes. As evident, this regime favors CHR emission from low-velocity charges. The relative enhancement, however, is determined by the ratio of the corresponding values of ${\tilde \eta}_{-1}$ 
at different velocities within a shared frequency range.
\begin{table}[h!]
	\centering
	\begin{tabular}{c c c c}
		\hline 
		\hline
		$\beta$ \quad & \quad $0.01$ \quad &\quad  $0.10$ \quad & \quad $0.25$ \quad \\
		$1/(\gamma-1) \quad $ & \quad $2 \times 10^4$ \quad &\quad $2 \times 10^2$ \quad & \quad 30 \quad \\
        $\omega_0 \,\,  [{\rm meV}]$  \quad & $0.5 \,  $ & $5.5 \, $ &\quad  $29 \, $    \\
        $\langle{\tilde \eta}_{-1}\rangle \,\, [{\rm Watt \, m}]^{-1}$ \quad & \quad $6\times10^{6}$ \quad & $6 \times 10^{4}$ \quad  \quad &  \quad $9 \times10^{3}$ \quad \\
        \hline
		\hline
	\end{tabular}
	\caption{The amplification factor $1/(\gamma-1)$   for different charge velocities, together with the corresponding maximum allowed frequency and the resulting value  of $\langle {\tilde \eta}_{-1}\rangle$ calculated at $\Omega =1$.}
	\label{T1}
\end{table}

\section{Summary and conclusions}

We consider the radiation of a charge $q$ moving at constant velocity $\mbf{v}= v \hat{\mbf {k}}$ in isotropic chiral matter with $\epsilon$, $\mu=1$ and magnetoelectric parameter $\sigma$.
By solving Maxwell's equations in the space-frequency domain using cylindrical coordinates, we obtain closed-form expressions for the electromagnetic fields 
(\ref{TOTALE}) and (\ref{TOTALB}).
 Each field is a superposition of two circular polarizations, $\nu= \pm $ , which represent  the normal modes of the system. 
 As a partial verification of our results we compare   with some explicit calculations in the iterative method proposed in Refs. \cite{Altschul:2014bba,Schober:2015rya, Altschul:2017xzx,DeCosta:2018nyf}. To this end, we take their expressions in the space-time domain, indicated in the second column of the Table \ref{tab:campos} and calculate the time-Fourier transform shown in the third column of the Table. Then we successfully compare these results with the expansion in powers of $\sigma$ of our exact expressions  (\ref{TOTALE}) and (\ref{TOTALB}), shown in the Appendix.

Although the  Poynting vector $\mbf{S}$ is not gauge invariant, we demonstrate that the total time-integrated energy flux over a closed surface at infinity preserves gauge invariance, defining the observable spectral energy distribution ${\cal E}$ indicated in Eq. (\ref{FINALSED})
Next, we consider the radiation zone by taking an infinite cylinder with its axis aligned with the particle motion and impose causality demanding outgoing waves as 
$\rho \to \infty$.  This condition defines the dispersion relation for each polarization mode, as shown in Eq. (\ref{POSDEF1}), determining the allowed frequency range for excitation given a specific set of parameters. 
The  explicit calculation of the spectral energy distribution in terms of the fields and potentials in the radiation zone shows that interference effects exactly cancel, yielding an independent superposition of the two polarization modes: ${\cal E}=
{\cal E}_+ + {\cal E}_-$.
Effectively, this means $\mbf{S}= \mbf{S}_+ + \mbf{S}_-$, such that our proof of gauge invariance remains valid for each polarization, providing independent physical significance to each mode. The spectral energy distribution for each polarization are given in alternative forms in Eqs (\ref{ELAMBDA2}) and (\ref{ELAMBDA}), as closed expressions in the parameters $\sigma$ and $v$.
 A further argument for the independent contribution of each polarization lies in their experimental distinguishability, marked by distinct emission angles that reflect the characteristic cones of CHR.
 We find at most two cones, determined by the angles (\ref{THETAC}), whose existence depends on fulfilling the relation (\ref{THETAC})— a condition that coincidentally matches the requirement for satisfying the dispersion relation (\ref{POSDEF1}).
The expressions for the spectral energy distribution of each mode in (\ref{ELAMBDA}) provide a direct proof that both ${\cal E}_\nu$ are positive, thus resolving for this particular case the problem posed by the fact that the local energy density is not positive definite in general.
Finally, we examine the frequency range for excitation of each polarization mode for a given set of parameters $n, \sigma, v$. 

Two cases deserve attention.
(i) Notably, when $v<c/n$, which is prohibited in the standard case, the mode $\nu=-$ is permitted within  the frequency range $0< \omega < \sigma \beta/(1-n^2\beta^2)$. In other words, this sector  of the model allows for the study of the most sought-after threshold-free CHR arising from slowly-moving charges  \cite{liu2017integrated, zhang2022tunable, hu2020nonlocality,  gong2023interfacial}. From   
 Eq.(\ref{FEE1}) together with the Fig. \ref{FIG11} and the Table \ref{T1} we find that the photon extraction efficiency depends crucially on the amplification factor $1/(1-\gamma)$  which increases significantly as the velocity decreases.
The specific scenario, where $n=1$ and $\sigma \neq 0$  is commonly referred to in the literature as "vacuum" Cherenkov radiation, which our calculation indicate is present within the specified frequency range. 
(ii) When $v> c/n$  and $\omega > \sigma \beta/(n^2 \beta^2-1)$ both polarization modes are excited. In other words, if we sweep our radiation detector across the entire angle, we are going to detect two well-defined radiation cones, each one radiating with ${\cal E}_\nu$, specifically depending on the parameter $\sigma$. However   the sum ${\cal E}_+  + {\cal E}_-$ results independent of $\sigma$ as shown in Eq. (\ref{ELAMBDA4}).  We interpret this property only  as a mathematical curiosity that in no way indicates some kind of cancellation between the radiation emitted by each cone, which otherwise occurs at different angles. The dependence on the velocity as well as on the parameter $\sigma$ of the emission angles for each mode are shown in Fig. \ref{FIG12} as a function of the frequency.  

A two-cone structure is also known to occur in bi-isotropic chiral media exhibiting optical activity, where the splitting arises from distinct refractive indices for circular polarizations~\cite{Bolotovskii1963,engheta1990vcerenkov}. While the physical origin in those cases is different, this analogy supports the experimental relevance of our predictions. It would be interesting to explore whether known modifications of radiation processes in optically active media, such as the fields of a uniformly moving charge, transition radiation, or Cherenkov radiation in chiral waveguides~\cite{barsukov1999vavilov,Galyamin2013Jul,Galyamin2017Mar}, also admit counterparts in isotropic chiral matter.

\section*{Appendix}

We present the expansion to order $\sigma^2$ of our general expressions for the electric and magnetic fields 
\beq
 E_\rho =\frac{2 q k^2}{\epsilon \omega} K_1(k\rho)+\frac{q \rho^2 \omega }{4c^4}\sigma^2 K_1(k\rho),\qquad \quad E_\phi=\frac{i q \rho \omega}{c^3} 
\sigma K_0(k\rho),\nonumber\\
\eeq
\beq
E_z=-\frac{2iqk^2}{\epsilon \omega}K_0(k\rho)-\frac{i q \rho v }{4c^4}\sigma^2\left(k\rho K_0(k\rho)-2 K_1(k\rho) \right),\notag \\
\eeq
\beq
B_\rho=-\frac{i q \rho k}{c^2}\sigma K_0(k\rho), \quad \qquad 
B_\phi=\frac{2kq}{c}K_1(k\rho)+\frac{q \rho}{c^3}\sigma^2 K_0(k\rho), \notag\\
\eeq
\beq
  B_z=\frac{q}{c^2}\sigma\left(2 K_0(k\rho)-k\rho K_1(k\rho) \right).
 \label{EXPSIGMA}
\eeq
Lack of space prevent us for showing the expansion for each polarization mode, which certainly  provides  a richer phenomenological information.

\section*{Acknowledgements}

R.M.v.D. and L.F.U. has been partially supported by DGAPA-UNAM Project No. AG100224 and by Project CONACyT (M\'{e}xico) No. 428214. R.M.vD. was supported by UNAM Posdoctoral Program (POSDOC). E.B.-A. and M.A.G. acknowledge support from  the Priority 2030 Federal Academic Leadership Program, and partial  support from the Foundation for the Advancement of Theoretical Physics and Mathematics ``Basis''. E.B.-A. acknowledges partial support from SECIHTI (México) doctoral scholarship. We thank Professor R. Potting for useful comments.

\bibliographystyle{elsarticle-num} 
\bibliography{references}
\end{document}